# Non-functional Requirements Documentation in Agile Software Development: Challenges and Solution Proposal

Woubshet Behutiye[1], Pertti Karhapää[1]

Dolors Costal[2], Markku Oivo[1] and Xavier Franch [2]

[1] University of Oulu, Pentti Kaiteran Katu 1, 90014 Oulu, Finland
[2] Universitat Politècnica de Catalunya, Campus Nord, Jordi Girona, 1-3, 08034 Barcelona, Spain
{woubshet.behutiye,Pertti.karhapaa,markku.oivo}@oulu.fi
{dolors,franch}@essi.upc.edu

**Abstract.** Non-functional requirements (NFRs) are determinant for the success of software projects. However, they are characterized as hard to define, and in agile software development (ASD), are often given less priority and usually not documented. In this paper, we present the findings of the documentation practices and challenges of NFRs in companies utilizing ASD and propose guidelines for enhancing NFRs documentation in ASD. We interviewed practitioners from four companies and identified that epics, features, user stories, acceptance criteria, Definition of Done (DoD), product and sprint backlogs are used for documenting NFRs. Wikis, word documents, mockups and spreadsheets are also used for documenting NFRs. In smaller companies, NFRs are communicated through white board and flip chart discussions and developers' tacit knowledge is prioritized over documentation. However, loss of traceability of NFRs, the difficulty in comprehending NFRs by new developers joining the team and limitations of documentation practices for NFRs are challenges in ASD. In this regard, we propose guidelines for documenting NFRs in ASD. The proposed guidelines consider the diversity of the NFRs to document and suggest different representation artefacts depending on the NFRs scope and level of detail. The representation artefacts suggested are among those currently used in ASD in order not to introduce new specific ones that might hamper actual adoption by practitioners.

## 1    Introduction

Non-functional requirements (NFRs) also referred to as quality requirements [21], represent software requirements that describe how software should perform [5]. These, for instance include software requirements about performance, usability, maintainability, reliability, and security. NFRs are characterized as vague and hard to define [17] and quite often result in being under/un-specified and undocumented. In particular, this is reflected in agile software development (ASD) where working software is prioritized over comprehensive documentation [2].

ASD's focus on "individuals and interaction over processes and tools" encourages minimal documentation [2]. ASD relies on tacit knowledge of the team and leans towards reducing the focus on requirements specification and documentation. Additionally, ASD is characterized with short iterations and it focuses on the quick delivery of working software. In such cases, developers face time pressure, mainly focus on delivery of functionalities and often do not give consideration to NFRs [6]. However, in such scenarios, neglecting NFRs may result in documentation debt with further consequences of increase in maintenance cost and effort [16].

NFRs play important role in the success of software systems [5, 9]. In ASD, existing requirements engineering practices fail short regarding the documentation of NFRs. For instance, user stories of ASD have limitations in specifying and documenting NFRs [15]. When NFRs are not documented, traceability becomes difficult, the likelihood of forgetting NFRs increases and consequences such as weak user acceptance may also result [7].

The findings from the scientific literature acknowledge the significance of handling NFRs in ASD [3, 8, 15]. The challenges of NFRs documentation in ASD, the limitations of ASD for handling NFRs, solution proposals for handling NFRs in ASD and the need for further investigation of the topic are reported frequently.

In this paper, we present the challenges of NFRs documentation in ASD and NFRs documentation practices identified from scientific literature and an ongoing empirical study in the Q-Rapids project [1][10], about managing NFRs in ASD. We also present guidelines for addressing challenges of NFRs documentation in ASD.

The rest of the paper is structured as follows. Section 2 describes the related work on challenges of documentation of NFRs and current ASD practices for documenting NFRs. Section 3 presents analysis of NFRs documentation practices and challenges identified from the ongoing empirical study about management of NFRs in ASD. Section 4 presents guidelines proposal for addressing documentation of NFRs in ASD. Finally, section 5 presents the conclusion.

## 2 Related work

### 2.1 Non-functional Requirements Documentation Challenges and Practices in Agile Software development

Research in the documentation and optimal integration of NFRs in ASD has paramount importance considering the vague nature of NFRs [17] and limitations in documentation practices of ASD [15]. Consequently there have been many studies investigating the topic area [8, 14, 15, 20]. In what follows, we present some challenges of NFRs management and current practices for documenting NFRs in ASD.

ASD puts less emphasis on the documentation of NFRs. Instead, its reliance on the continuous interaction with customers is thought to minimize the need for specifying

---

[1] http://q-rapids.eu/

NFRs [20]. In ASD, NFRs are ill defined and rarely documented, and there are no formal acceptance tests for NFRs. As a result, problems arise at later stages of development [14].

The negligence of NFRs appears to be a major concern of many agile projects and is reported frequently [4, 14, 17]. For instance, Cao and Ramesh [4] identified the neglect of NFRs and minimal documentation as major challenges of agile requirements engineering in an empirical investigation of 16 software development organizations. According to their findings, NFRs are given less priority in the early stage of ASD as customers instead prioritize core functionality. Consequently, minimal documentation and negligence of NFRs in ASD result in challenges of scalability of the software, and introduce difficulty for new members joining the development team.

Failure to consider NFRs in the early stages of software development may result in poor quality software, increased maintenance costs and time [5]. Indeed, when NFRs are omitted in the early stages of development, they result in major issues at later stages. ASD methods face challenges in addressing specific NFRs such as security [1]. For instance, Scrum's lack of consideration for integrating security (NFRs) in the development process opens vulnerability to the software [1]. Absence of documentation for security, limited amount of time for testing security in sprints, and difficulty for integrating security related activities are major security issues in Scrum.

ASD mainly utilizes index cards, paper prototypes and storyboards to document features and requirements [14]. Practices such as user stories are used for documenting high level requirements [4]. However, they have limitations for specifying and documenting NFRs [11, 12, 15]. Martakis et al. [15], found that agile developers face challenges while using user stories for documenting NFRs such as security and internationalization.

Consequently, there have been proposals for integrating, planning and managing NFRs in ASD (e.g. AFFINE framework, NORMAP, NORPLAN, security backlog for Scrum etc.) [3, 8, 15]. Lightweight practices and systematic solutions that integrate NFRs in ASD without compromising quality of software and agility of the development process are of high importance.

## 3 Non-functional Requirements Documentation Practices and Challenges in ASD Projects

We conducted case studies following [19], in four case companies that are part of the Q-Rapids project, in order to synthesize knowledge regarding management of NFRs in ASD. We collected data through semi-structured interviews and applied qualitative analysis on the transcriptions of the interviews. The four case companies providing the use cases (UCs) for the project are of varying size and domain. The first company has over 900 employees while the second has over 600 employees. The third is large scale global company with over 100,000 employees while the fourth has less than 100 employees. We conducted 12 interviews, with roles that include product owners, project managers, developers and quality assurance engineers, DevOps Specialist, and Scrum masters.

Agile practices and iterative development are applied in all the UCs, of which three are close to Scrum. In UC1, the company follows in-house tailored agile and iterative development. However, they do not have any fixed sprint cycles. In comparison, the development applied in UC2 and UC4 is the closest to Scrum with daily sprints and weekly, or biweekly sprints. In UC3, which is the large-scale company, Scrum, or variations of it, is applied in some of the development teams at lower levels of the organization. In UC3, a team can apply any development model they see fit. Continuous integration is applied in all the UCs.

The interview findings reveal that the companies employ varying practices for documenting both functional requirements (FRs) and NFRs. UC1 prefers to focus effort on development and documents requirements in detail only when implementing features that the developers are unfamiliar with. NFRs are communicated through whiteboards during meetings. On the other hand, UC2 and UC3 document both FRs and NFRs. Partly this is enforced through standards that the companies must comply with. In UC2 requirements are documented in epics, features, and user stories, and NFRs are also in the acceptance criteria and Definition of Done (DoD). Additionally, word documents, PowerPoints and wikis are used for documentation during the development. Along the process, the documentation in the wikis becomes more of a technical description of the software and the connection to the original high level requirements is lost. The interviewees suggested including more design documentation in the user stories to preserve this link. Using Word and PowerPoint for documentation is perceived challenging, as these documents become easily detached from the actual software. This is due to the fact that it is easy to forget updating a certain document with every change to the code.

In the case of UC3, which is a large and distributed organization, documentation is important as there are teams in different locations that may be working on the same feature. There is complex backlog structure and all the requirements are also documented in features that are broken down into sub features and further into tasks that can be coded. Additionally, NFRs are documented in DoD and acceptance criteria. At the lower task level, however, there are no NFRs in the backlog as such, but the tasks need to meet the DoD including quality criteria. In UC3, documentation of NFRs is identified as problematic. Our interviewees find the requirements management tool under use and complexity of backlogs difficult and stated that they are not able to identify dependent NFRs. Additionally, internally inherited NFRs such as operability are rarely documented and prioritized. UC4 documents all the requirements (FRs and NFRs) in the epics and user stories. DoD and acceptance criteria (at user story, task and ticket levels) are used for documenting NFRs. Additionally, excel spread sheets, mock-ups, product backlogs and sprint backlogs are used for documenting NFRs.

In summary, we observe that three of the UCs follow up procedures for documenting NFRs in ASD. The UCs followed a formal approach to specify and document NFRs. However, in one UC, NFRs were not documented and were rather communicated in face-to-face meetings facilitated by whiteboards and flip charts. In such cases, companies relied on the tacit knowledge of the developers. These developers discuss NFRs in meetings (e.g. daily stand-ups, sprint planning meetings) and avoid detailed documentations. **Table 1** summarizes NFRs documentation practices and challenges identified from the UCs.

5**Table 1.** Summary of NFRs documentation practices and challenges in ASD UC companies

| Use case | NFRs documentation practice | NFRs documentation challenge |
|---|---|---|
| UC1 | NFRs are not formally documented, however communicated through white board and when necessary documented in word documents | NFRs not documented properly and resulted in the lack of traceability of NFRs, difficulty for new developers joining team |
| UC2 | NFRs documented in epics, features, and user stories, acceptance criteria and DoDs, wiki pages, word docs with FRS | Lower-level details are lost in documentation, word and power point documents disconnected from actual software |
| UC3 | NFRs documented in features, acceptance criteria and DoDs in complex backlogs | Complexity of backlogs makes it hard to identify dependent NFRs, internally generated NFRs are not documented |
| UC4 | NFRs documented in epics, user stories, in DoD and acceptance criteria (at user story, task and ticket levels), in product and sprint backlogs. Mockups, wireframes, word, spreadsheet are also used for documenting NFRs while Whiteboards and flip charts facilitate communication of NFRs. | Not reported by interviewees |

Our findings reveal that companies may face challenges when they fail to document NFRs properly. For instance, in UC1 when relying on tacit knowledge of developers', the traceability of NFRs becomes difficult in later stages of development. The interviewees pointed out that this introduces challenge to new developers joining the team as they will have limited visibility of the NFRs. Scientific literature depicts similar findings [11]. On the other hand, difficulty in identifying interdependent NFRs in complex backlogs is another challenge identified in UC3.

The significance of NFRs for the success of software projects and specific challenge of ASD in documenting NFRs that is also reflected in the UCs, prompt us to propose lightweight and systematic guidelines for documenting NFRs in ASD.

## 4  Guidelines Proposal for Documenting NFRs in ASD

In order to cope with the diversity of approaches to represent requirements in agile methods, we take the following assumptions that do not compromise the general applicability of our approach: 1) FRs are specified using both epics and user stories, 2) user stories may include one or more acceptance criteria and 3) user stories will be derived from epics and this link will be recorded.

The system NFRs to document may be quite diverse. Remarkably the scope of NFRs may vary significantly. A NFR may refer to quality properties of the entire system to be developed but it also may define quality properties for a particular service, function or system component [18]. We distinguish three different types of scope for NFRs:

6system-wide for those that apply to the entire system, group-wide for those that apply to a set of user stories (or a group of functionalities) and local for those that apply to a single user story (or functionality). Additionally, the level of detail in which a NFR is specified may vary. Accordingly, we distinguish among generic NFRs, i.e., specified at a high level of abstraction (near to the notion of goal) [13], and detailed NFRs, i.e., specified as a concrete feature or tied to a concrete solution. Quite often, a generic NFR may be specified in an earlier development stage and, later on, it may be refined into a set of detailed NFRs that operationalize it (e.g. the generic NFR "The system must be usable" may be refined into "The system must allow reaching any functionality in no more than 3 clicks" among other detailed NFRs). All combinations of scope and detail are possible when specifying a NFR. For instance, "The critical functions of the system must take less than 0.25 seconds, 90% of the times" is group-wide and detailed while "The functionality for checking the account balance must have a good response time" is local and generic.

The variability of NFRs both in scope and detail suggests that there is not a single representation artefact that is adequate to cope with all of them. Therefore, a proposal for documenting NFRs in ASD should provide different artefacts for representing them and a set of guidelines to select the most adequate representation depending on the features of each specific requirement. In our opinion, the artefacts should preferably be those currently used in ASD in order not to introduce new specific artefacts that might damage the agility of the process and hamper actual adoption by practitioners. Therefore, our guidelines proposal, summarized in **Table 2**, consists of using either acceptance criteria, user stories or epics to represent NFRs.

**Table 2.** Guidelines for documenting NFRs according to their scope and detail

| Scope | Detail | Representation artefact | Observation |
|---|---|---|---|
| Local | Generic | User story (NFR user story) | With a link to the functional user story to which it applies |
| | Detailed | Acceptance criteria | Appearing in the functional user story to which it applies |
| Group wide | Generic | Epic | The description of the epic must clarify to which group of functionalities it applies (e.g. "critical functions of the system") |
| | Detailed | (1) User story or (2) Acceptance criteria | (1) The description of the user story must clarify to which group of functionalities it applies or include links to the user stories it applies (2) Appearing in the functional user stories to which it applies |
| System wide | Generic | Epic | The description of the epic must clarify it is system-wide (e.g. by referring to "the system") |
| | Detailed | User story | The description of the epic must clarify to which group of functionalities it applies (e.g. "critical functions of the system") |

In the following, we describe the rationale used in our proposal (see **Table 2**) to select the adequate representation artefact for a NFR based on the scope and detail of the NFR.

The simplest case is that of local and detailed NFRs. They can be locally represented, in the affected user story, as acceptance criteria, because these NFRs neither affect the other user stories nor need further refinements. Conversely, local and generic NFRs cannot be documented as acceptance criteria because they are not concrete enough. Therefore we propose to document them as user stories that should be linked to the functional user story to which they apply. Then, the acceptance criteria of this latter user story may refine the generic NFR.

For system-wide NFRs, we propose to use epics if they are generic and user stories if they are detailed. System-wide and generic NFRs are documented by epics because they are high level qualities of the whole system and thus they are relevant requirements that will probably need to be further detailed by means of user stories (derived from that epic). These latter user stories will then be representing system-wide and detailed NFRs.

For group-wide NFRs, our proposal is similar to that of system-wide NFRs. However, if they are detailed and the group of functionalities affected by the NFRs is small, we propose, as an additional option to document them as acceptance criteria of the user stories to which they apply (like local and detailed NFRs).

## 5    Conclusion

In this paper, we presented the findings of NFRs documentation practices in ASD projects. We identified that NFRs are documented together with FRs. The UCs applied epics, features, user stories, acceptance criteria and DoD of user stories, and backlogs to document NFRs. Whiteboard and flip charts are used to facilitate the communication of NFRs in cases where they are not documented. The difficulty in the traceability of NFRs, problems in identifying interdependent NFRs and detached documentation from actual software, were among the challenges of NFRs identified in the UCs. Moreover, we propose guidelines for documenting NFRs in ASD. The proposed guidelines acknowledge diversity of NFRs and utilize existing ASD artefacts such as epics, user stories and acceptance criteria for documenting NFRs. In addition, the guidelines consider different levels for the scope and details of abstraction of NFRs.

**Acknowledgments**. This work is a result of the Q-Rapids project, which has received funding from the European Union's Horizon 2020 research and innovation program under grant agreement N° 732253.

8